\documentclass{article}
\usepackage{spconf,amsmath,graphicx}
\usepackage{color}
\usepackage[ruled,linesnumbered]{algorithm2e}
\usepackage{listings}
\usepackage{booktabs}
\usepackage{multirow}
\usepackage{pifont}
\PassOptionsToPackage{hyphens}{url}\usepackage{hyperref}

\title{Fast-U2++: Fast and Accurate End-to-End Speech Recognition in Joint CTC/Attention Frames}
\name{\begin{tabular}{c}
     Chengdong Liang$^{1,2,3}$, Xiao-Lei Zhang$^{1,*}$, BinBin Zhang$^{2,3}$, Di Wu$^{2,3}$, Shengqiang Li$^{2,3}$, \\Xingchen Song$^{2,3}$, Zhendong Peng$^{2,3}$, Fuping Pan$^2$
\end{tabular}
}
\address{
	$^1$School of Marine Science and Technology, Northwestern Polytechnical University, Xi'an, China\\
	$^2$Horizon Robotics, Beijing, China~
	$^3$WeNet Open Source Community \\
	$\text{\small liangchengdong@mail.nwpu.edu.cn, xiaolei.zhang@nwpu.edu.cn}$
	\thanks{* Corresponding author.}
	}

\begin{document}
%
\maketitle
\begin{abstract}
Recently, the unified streaming and non-streaming two-pass (U2/U2++) end-to-end model for speech recognition has shown great performance in terms of streaming capability, accuracy and latency. In this paper, we present fast-U2++, an enhanced version of U2++ to further reduce \textit{partial latency}. The core idea of fast-U2++ is to output partial results of the bottom layers in its encoder with a small chunk, while using a large chunk in the top layers of its encoder to compensate the performance degradation caused by the small chunk.
Moreover, we use \textit{knowledge distillation} method to reduce the token emission latency. We present extensive experiments on Aishell-1 dataset. Experiments and ablation studies show that compared to U2++, fast-U2++ reduces \textit{model latency} from 320ms to 80ms, and achieves a character error rate (CER) of 5.06\% with a streaming setup.

\end{abstract}

\section{Introduction}
\label{sec:intro}
In recent years, end-to-end (E2E) automatic speech recognition (ASR) received more and more attention. Compared with conventional hybrid DNN-HMM systems \cite{li2020developing,sainath2020streaming}, E2E models not only have a simple training and decoding process, but also have better performance than the hybrid systems. The most popular E2E models are based on the connectionist temporal classification (CTC) \cite{graves2006connectionist,amodei2016deep}, recurrent neural network transducer (RNN-T) \cite{graves2012sequence,graves2013speech}, and attention-based encoder-decoder (AED) \cite{chan2016listen,chorowski2015attention}.
The hybrid CTC/attention end-to-end ASR \cite{kim2017joint} jointly minimizes the CTC and attention decoder losses, which results in faster convergence and also improves the robustness of the AED model. During the decoding, it combines the attention score and the CTC score, and performs joint decoding.


However, deploying the E2E system is not easy, and there are many practical problems to be solved \cite{wu2021u2++}. The first challenge is the streaming problem. Some state-of-the-art models such as Transformer \cite{dong2018speech} and Conformer \cite{gulati2020conformer} could not run in streaming mode, which limits application scenarios. Second, streaming and non-streaming systems are usually developed, trained and deployed separately, and a lot of engineer efforts are required to promote E2E models to the production level.

To address the aforementioned problems, many efforts have been made. Specifically, for a unified non-streaming and streaming model, \cite{yu2020dual} proposed a dual-mode E2E for both streaming and non-streaming. It also uses knowledge distillation to train the student streaming mode from the full-context non-streaming mode. \cite{narayanan2021cascaded} proposed {cascaded encoders} which consists of both streaming and non-streaming encoders based on RNN-T. It first deals with input features by the streaming encoder, and then operates exclusively on the output of the streaming encoder by the non-streaming encoder. \cite{moritz2021dual} proposed the dual causal/non-causal self-attention for E2E ASR. We have also proposed a unified streaming and non-streaming E2E model (U2/U2++) \cite{zhang2020unified,wu2021u2++}, and opened a production first and production-ready E2E toolkit, named WeNet, \cite{yao2021wenet,zhang2022wenet} for the aforementioned challenges in a simple and elegant way.

Although many efforts have been made to U2/U2++ and WeNet, low latency is always desired in real-world applications for a good customer experience.
In this paper, we focus on optimizing \textit{partial latency} of U2++, where the {partial latency} is defined as the time delay from the time when a user finishes speaking and the time when the first correct partial hypothesis is generated by the model. The {partial latency} is affected by (i) the \textit{model latency} which is the waiting time introduced by the model structure, and (ii) the token emission latency \cite{shangguan2021dissecting}. For the U2++ model, we found that the CTC spike of the streaming model falls far behind that of the non-streaming model. This problem is mainly caused by causal convolution.


Based on the above analysis, in this paper, we propose fast-U2++ to further reduce the latency of U2++. The contributions are summarized as follows:
\begin{itemize}
    \item First, we propose a dual-mode Conformer encoder. Compared with the conventional Conformer encoder \cite{gulati2020conformer}, we replace the conventional non-causal convolution with the dual-mode convolution. We demonstrate that the rescoring prediction of the steaming mode significantly benefits from the \textit{joint training} of its non-streaming mode.
    \item Second, we use a small chunk in the bottom layers of the encoder to output partial results, which significantly reduces the \textit{model latency}.
    \item Moreover, we reduce the token emission latency of the streaming mode by shifting the prediction of the non-steaming mode by \textit{knowledge distillation} \cite{hinton2015distilling,yu2020dual}.
\end{itemize}
With the above improvement, fast-U2++ can output partial results from the bottom layers of encoder efficiently, and can also dynamically control the trade-off between the latency and the performance, when working in the streaming mode.


\begin{figure}[t]
	\centering
	\includegraphics[width=\linewidth]{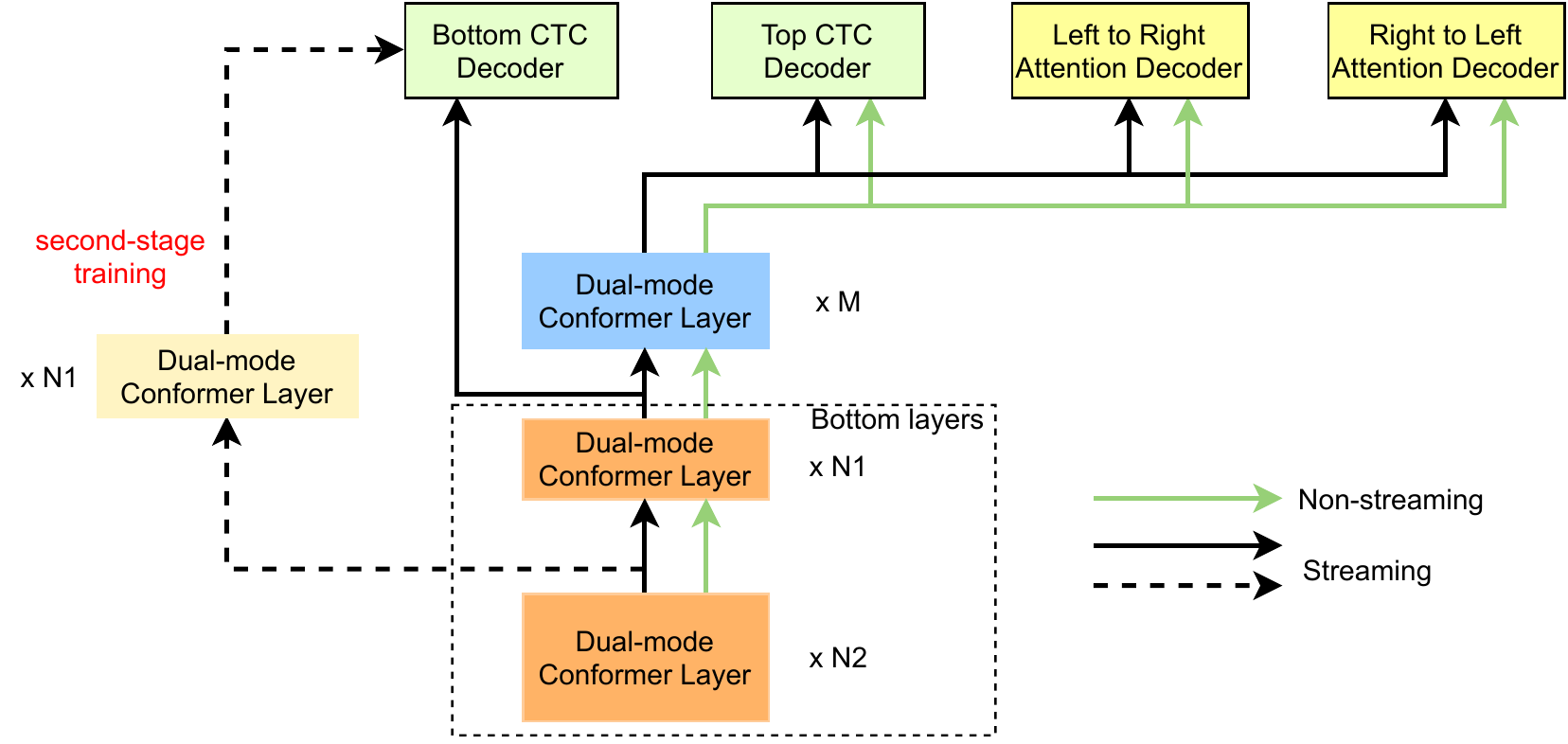}
	\caption{Architecture of the proposed fast-U2++.}
	\label{fig:fastu2pp}
\end{figure}

%
%

\section{fast-U2++}
\label{sec:pagestyle}

\subsection{Model architecture}
The proposed model architecture is shown in Figure~\ref{fig:fastu2pp}. It contains four parts: a shared dual-mode encoder that models the context of the input acoustic features, a CTC decoder that models the alignment of the frames and tokens, a Left-to-Right (L2R) attention decoder that models the dependency of the left tokens, and a Right-to-Left (R2L) attention decoder that models the dependency of the right tokens.

The shared dual-mode encoder consists of multiple dual-mode Conformer layers. The dual-mode Conformer layer uses a dual-mode convolution. The CTC decoder consists of a linear layer and a log-softmax layer. It is shared by the streaming output of the bottom layers of the encoder and the non-streaming output of the top layers of the encoder.
The L2R and R2L attention decoder use the conventional Transformer decoder.

\begin{figure}[t]
	\centering
	\includegraphics[width=0.7\linewidth]{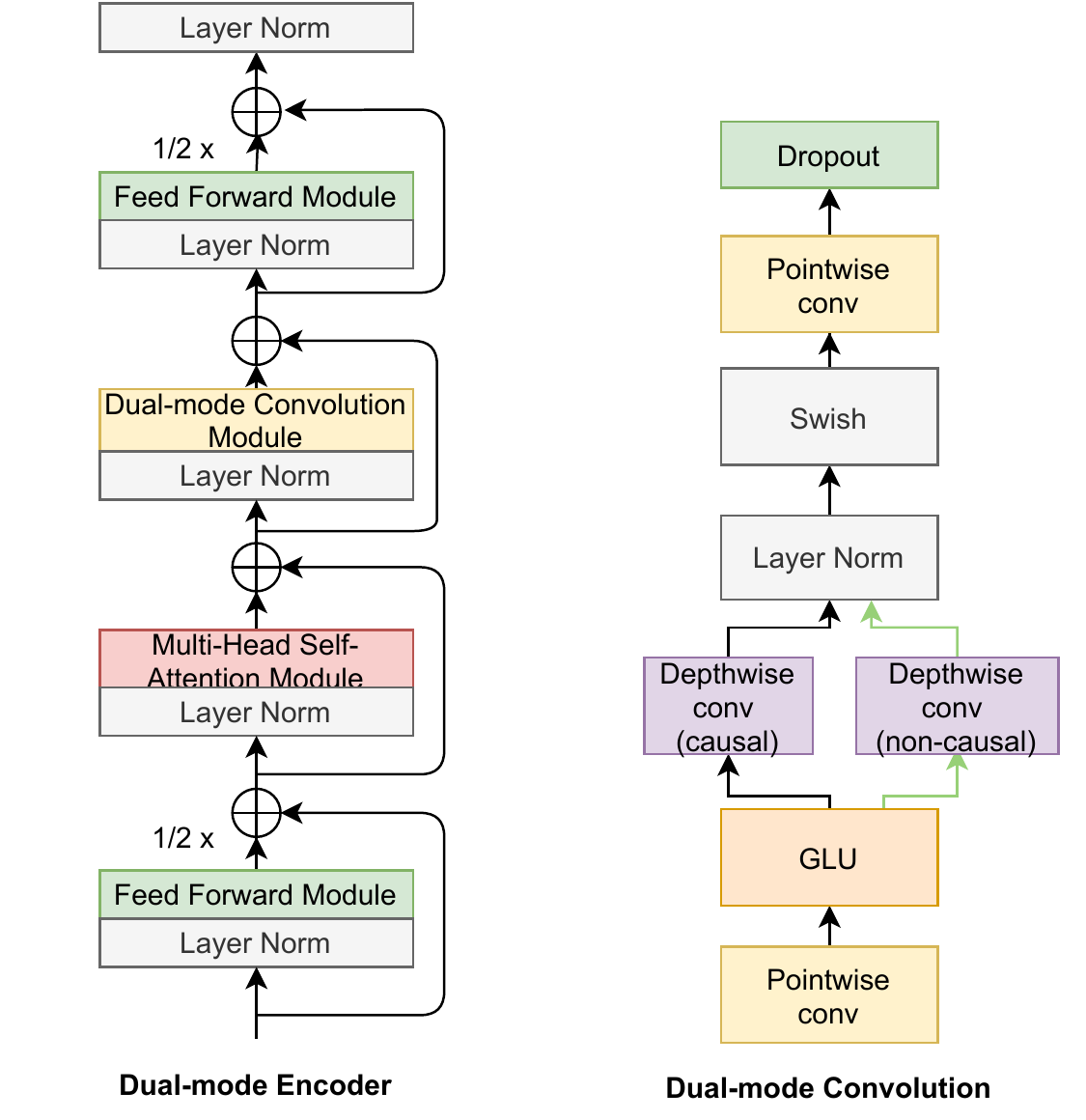}
	\caption{Architecture of the proposed dual-mode encoder.}
	\label{fig:encoder}
\end{figure}

\subsubsection{Dual-mode convolution}
\label{sec:dual-conv}
Because the convolution layer of the conventional Conformer considers both left and right context, it cannot be used in the streaming mode.
To overcome this problem, as shown in Figure~\ref{fig:encoder}, fast-U2++ adds casual convolution \cite{tripathi2020transformer} for the streaming mode, as a parallel branch of the conventional convolution for the non-streaming mode. Because both the casual convolution and the conventional convolution utilize the depthwise convolution structure which contains few parameters, while the other parts of the streaming and non-streaming modes are shared, the increased parameters induced by the casual convolution can be ignored.

\subsection{Training}
In this section, we discuss two important training techniques: \textit{joint training} and \textit{knowledge distillation}.
\subsubsection{Joint training}
Suppose $\mathbf{x}_{\mathrm{s,l}}$, $\mathbf{x}_{\mathrm{s,h}}$ and $\mathbf{x}_{\mathrm{ns,h}}$ are the output of bottom layers in the encoder with the streaming mode, output of the top layers in the encoder with the streaming mode, and output of the top layers in encoder with the non-streaming mode, respectively. Let $\mathbf{y}$ denote the corresponding text. The training loss $\mathbf{L}_{\mathrm{joint}}(\cdot)$ for the joint training of streaming and non-streaming mode is a combination of the streaming ASR loss of the bottom layers, streaming ASR loss of the top layers,  and non-streaming ASR loss of the top layers:
\begin{equation}
    \label{eq:train}
    \mathbf{L}_{\mathrm{joint}} = \mathbf{L}_{\mathrm{asr}} \left(\mathbf{x}_{\mathrm{s,l}},\mathbf{y}\right) + \mathbf{L}_{\mathrm{asr}} \left(\mathbf{x}_{\mathrm{s,h}},\mathbf{y}\right) +  \mathbf{L}_{\mathrm{asr}} \left(\mathbf{x}_{\mathrm{ns,h}},\mathbf{y}\right)
\end{equation}
where $\mathbf{L}_{\mathrm{asr}}(\cdot)$ is the combined CTC and AED loss for ASR:
 \begin{equation}
    \label{eq:asr}
    \mathbf{L}_{\mathrm{asr}} \left(\mathbf{x} ,\mathbf{y}\right) = \lambda \mathbf{L}_{\mathrm{CTC}}\left(\mathbf{x}, \mathbf{y}\right) + \left(1 - \lambda\right) \mathbf{L}_{\mathrm{AED}} \left(\mathbf{x}, \mathbf{y}\right)
\end{equation}
where $\lambda$ is the weight of the CTC loss, and the AED loss $\mathbf{L}_{\mathrm{AED}}(\cdot)$ consists of a L2R AED loss $\mathbf{L}_{\mathrm{L2R}}(\cdot)$ and a R2L AED loss $\mathbf{L}_{\mathrm{R2L}}(\cdot)$:
\begin{equation}
    \label{eq:aed}
    \mathbf{L}_{\mathrm{AED}} \left(\mathbf{x}, \mathbf{y}\right) = \left(1 - \alpha \right) \mathbf{L}_{\mathrm{L2R}} \left(\mathbf{x}, \mathbf{y}\right) + \alpha \mathbf{L}_{\mathrm{R2L}} \left(\mathbf{x}, \mathbf{y}\right)
\end{equation}
with $\alpha$ as the weight of R2L AED loss.

\subsubsection{Knowledge distillation}
\label{sec:distill}
Because the prediction of the non-streaming mode usually has lower latency than the streaming mode, we can control the token emission latency of the streaming mode by shifting the prediction of the non-streaming mode in the training stage. Based on this observation, we can compute a distillation loss from $\mathbf{x}_{\mathrm{s,l}}$ and $\mathbf{x}_{\mathrm{ns,h}}$ by:
\begin{equation}
    \label{eq:distill}
     \mathbf{L}_\mathrm{distill} = \operatorname{SmoothL1Loss} \left(\mathbf{x}_{\mathrm{s,l}}, \mathbf{x}_{\mathrm{ns,h}}\right) / N_{\mathrm{frame}}
\end{equation}
Where $\operatorname{SmoothL1Loss}(\cdot)$ is the smoothed L1 loss described in PyTorch\footnote{https://pytorch.org/docs/stable/generated/torch.nn.SmoothL1Loss.html}, and $N_{\mathrm{frame}}$ is the number of speech frames.

We propose two different distillation methods:
 \begin{itemize}
 \item \textbf{D1}: Distill the last layer of the top layers in encoder and the bottom layers in encoder.
 \item \textbf{D2}: Distill the last two layers of the top layers in encoder and the bottom layers in encoder.
 \end{itemize}

\subsubsection{Training strategy}
We propose a two-stage training strategy. In the first stage, we only conduct the joint training. In the second stage, we add $N1$ dual-mode Conformer layers in the bottom layers of the encoder as shown in Figure~\ref{fig:fastu2pp}. Then, we only train the newly added Conformer layers and the corresponding CTC decoder by:
\begin{equation}
    \label{eq:fine}
    \mathbf{L} = \mathbf{L}_{\mathrm{joint}} + \beta \mathbf{L}_\mathrm{distill}
\end{equation}
where $\beta$ is the distillation weight.

\subsection{Decoding}
We use a similar two-pass rescoring decoding strategy with U2++. In the first pass, the bottom CTC decoder outputs the one-best candidates using greedy search decoding. At the same time, the top CTC deocder outputs n-best candidates using prefix beam search decoding.

In the second pass, the two attention decoder scores these n-best candidates. The final score is calculated by:
\begin{equation}
    \label{eq:score}
    \mathbf{S}_{\mathrm{final}} = \lambda  \mathbf{S}_{\mathrm{CTC}} + \left(1 - \alpha\right)  \mathbf{S}_{\mathrm{L2R}} + \alpha \mathbf{S}_{\mathrm{R2L}}
\end{equation}
The candidate with the best score is chosen as the decoding result finally.

\section{Experiments}
\label{sec:exp}
To evaluate the proposed fast-U2++ model, we carried out our experiments on the Chinese Mandarin speech corpus AISHELL-1\footnote{http://openslr.org/33/}. We used \textit{WeNet\footnote{https://github.com/wenet-e2e/wenet}} toolkit for all experiments.

We used the state-of-the-art ASR network U2++ Conformer as our baseline model. For the proposed fast-U2++, the parameter settings of the encoder and decoder were the same as U2++. The parameters $N1$, $N2$, $M$ were set to 2, 5, 5, respectively.
The input features are 80-dimensional log Mel-filter banks (FBANK) with a 25ms window and a 10ms shift. SpecAugment \cite{park2019specaugment} and SpecSub \cite{wu2021u2++} were applied in the same way as U2++. We set the label smoothing weight to 0.01. $\lambda$ was set to 0.3 for the joint training with the Cross-entropy loss. $\alpha$ was set to 0.3. Adam optimizer was used, and the learning rate was warmed up with 25,000 steps. The final model averages the top-30 models with the best validation loss on the development set. The beam size of the CTC prefix search was set to 10. We used the same dynamic chunk training as U2++ for the streaming mode, which enables the model to work well in different chunk sizes. Speed perturbation with 0.9, 1.0 or 1.1 was applied on the fly.
\begin{figure}[t]
	\centering	\includegraphics[width=\linewidth]{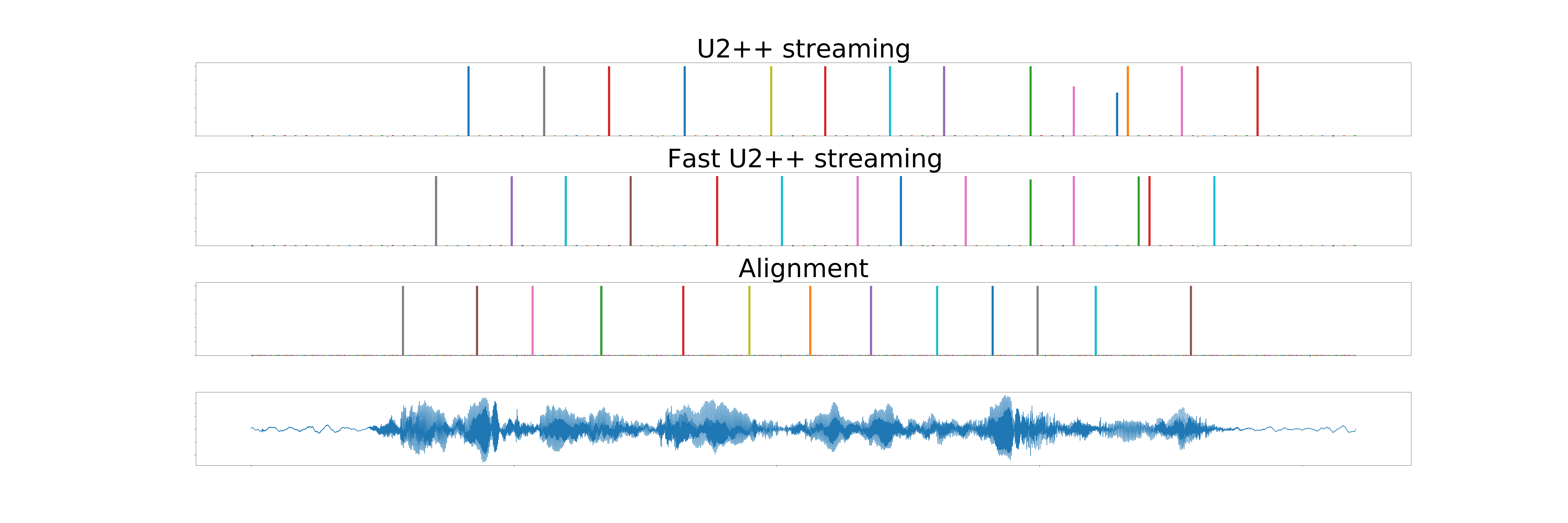}
	\caption{Comparison of the CTC spike distributions between U2++ and the proposed fast-U2++.}
	\label{fig:align}
\end{figure}

\textbf{Latency Metrics}. Motivated by \cite{shangguan2021dissecting,chang2020low}, the latency metrics are \textit{First Token emission Delay} (FTD) and \textit{Last Token emission Delay} (LTD). FTD is defined as the timestamps difference of the following two events: (1) the first token is emitted, (2) the first token is estimated. LTD is defined as the emission latency for the final token. For all experiments, we reported both 50-th (P50) and 90-th (P90) percentile values of all utterance excluding abnormal sentences, so as to better reflect the \textit{model latency}. We used a GMM/HMM model trained by the Kaldi\footnote{https://github.com/kaldi-asr/kaldi} toolkit to get the estimated timestamps.

\subsection{Main results}
Table~\ref{table:res} reports CER, FTD and LTD of the proposed fast-U2++ in the streaming setting on the AISHELL-1 dataset.
From Table~\ref{table:res}, we see that, with the \textit{knowledge distillation}, the CER of the CTC greedy search in the bottom layers is 7.34$\%$ when the chunk size is 4. To compensate for the performance degradation caused by small chunk size of the bottom layers, we set the chunk size of top layers to 24. Finally, the CER of rescoring is 5.06\%, which is close to the performance of the U2++ with chunk size of 16. For our chunk based decoding , the average waiting time is half of the chunk theoretically \cite{yao2021wenet}. Therefore, the \textit{model latency} of fast-U2++ is 80ms when chunk size of the bottom layers is 4. An example is given in Figure~\ref{fig:align} where the CTC spike of fast-U2++ shifted significantly forward.

Moreover, we studied the effect of the weight $\beta$ by setting it to 0.02, 0.03, and 0.05 respectively, together with two distillation methods, i.e. \textbf{D1} and \textbf{D2} as described in Section \ref{sec:distill}. As listed in Table~\ref{table:res}, when $\beta$ is 0.05, \textbf{D1} achieves better performance than \textbf{D2}. However, when $\beta$ is getting small, \textbf{D2} tends to achieve better performance than \textbf{D1}. Both methods significantly reduce FTD and LTD, where the improvement of LTD is more apparent than FTD.

Table~\ref{table:comp} shows the comparison results of the proposed method with other representative systems on the AISHELL-1 dataset, where $\Delta$ usually is about 100ms. From the table, we see that proposed fast-U2++ could not only have lower model latency, but also ensure good performance of final results.

\begin{table}[t]
\centering
\caption{CER, FTD and LTD of fast-U2++ on the test set of AISHELL-1. For U2++, the result of the greedy search decoding is obtained using the full 12-layers encoder. For fast-U2++, we calculate the result of the greedy search decoding with the bottom layers of its encoder. `4/24' means that the chunk size of the bottom layers in the encoder is 4, and the chunk size of the top layers in the encoder is 24. `16' means that the chunk size of the full 12-layers in the encoder is 16. }
\label{table:res}
\scalebox{0.77}{
\begin{tabular}{ccccclcl}
\toprule
\multirow{2}{*}{method}           & \multirow{2}{*}{\begin{tabular}[c]{@{}c@{}}decoding\\chunk size\end{tabular}} & \multicolumn{2}{c}{CER (\%)}        & \multicolumn{2}{c}{FTD (ms)} & \multicolumn{2}{c}{LTD (ms)}  \\
                                  &                                                                               & greedy & rescoring             & P50 & P90                    & P50 & P90                     \\
\midrule
\multicolumn{1}{l}{U2++ baseline} & 16                                                                            & 5.81   & 5.05                  & 248 & 300                    & 216 & 250                     \\
\midrule
\multicolumn{1}{l}{Fast-U2++}     & 4/24                                                                          & 7.47   & 5.08                  & 230 & 280                    & 160 & 210                     \\
\midrule
~ + D1 $\beta$=0.05               & \multirow{7}{*}{4/24}                                                         & 8.19   & \multirow{7}{*}{5.06} & 110 & 170                    & 40  & 90                      \\
~ + D1 $\beta$=0.03               &                                                                               & 7.82   &                       & 140 & 200                    & 50  & 100                     \\
~ + D1 $\beta$=0.02               &                                                                               & 7.65   &                       & 170 & 220                    & 60  & 110                     \\
~ + D2 $\beta$=0.05               &                                                                               & 8.27   &                       & 110 & 170                    & 40  & 80                      \\
~ + D2 $\beta$=0.03               &                                                                               & 7.77   &                       & 140 & 210                    & 50  & 90                      \\
~ + D2 $\beta$=0.02               &                                                                               & 7.56   &                       & 160 & 220                    & 50  & 110                     \\
~ + D2 $\beta$=0.01               &                                                                               & 7.34   &                       & 170 & 220                    & 70  & 120                     \\
\bottomrule
\end{tabular}}
\end{table}

\begin{table}[t]
\centering
\caption{CER results of fast-U2++ and the state-of-the-art streaming methods.}
\label{table:comp}
\scalebox{0.77}{
\begin{tabular}{llcc}
\toprule
Method               & Latency (ms) & LM & CER (\%)   \\
\midrule
Sync-Transformer \cite{tian2020synchronous}  & 400          & \ding{56}  & 8.91  \\
SCAMA \cite{zhang2020streaming}             & 600          & \ding{56}  & 7.39  \\
MMA \cite{inaguma2020enhancing}               & 640          & \ding{56}  & 6.60  \\
U2 \cite{zhang2020unified}                & 320 + $\Delta$         & \ding{56}  & 5.33  \\
WNARS \cite{wang2021wnars}               & 640          & \ding{52}  & 5.15  \\
U2++ \cite{wu2021u2++}              & 320 + $\Delta$          & \ding{56}  & 5.05  \\
Fast-U2++ (proposed) &  80 + $\Delta$            & \ding{56}  & 5.06  \\
\bottomrule
\end{tabular}}
\end{table}

\begin{table}
\centering
\caption{Ablation experiments without \textit{knowledge distillation}. `7-layers' means that the result of the greedy search decoding is obtained with the bottom layers of the encoder. The number of the bottom layers in the encoder is $7$ ($N1+N2 = 7$). `12-layers' means that the result of the greedy search decoding is obtained with the full 12-layers encoder.}
\label{table:ctc}
\scalebox{0.75}{
\begin{tabular}{cccccc} 
\toprule
\multirow{3}{*}{\begin{tabular}[c]{@{}c@{}}non-streaming\\loss\end{tabular}} & \multirow{3}{*}{\begin{tabular}[c]{@{}c@{}}CTC \\decoder\end{tabular}} & \multirow{3}{*}{\begin{tabular}[c]{@{}c@{}}decoding\\chunk size\end{tabular}} & \multicolumn{3}{c}{CER (\%)}                                  \\
                                                                             &                                                                        &                                                                               & \multicolumn{2}{c}{greedy} & \multirow{2}{*}{rescoring}  \\
                                                                             &                                                                        &                                                                               & 7-layers & 12-layers       &                             \\ 
\midrule
\ding{56}                                                                           & no shared                                                              & 4/24                                                                          & 7.49     & 5.99            & 5.26                        \\
                                                                             &                                                                        & 16                                                                            & -        & 5.84            & 5.23                        \\
\ding{56}                                                                           & shared                                                                 & 4/24                                                                          & 7.46     & 5.82            & 5.34                        \\
                                                                             &                                                                        & 16                                                                            & -        & 5.73            & 5.24                        \\ 
\midrule
\ding{52}                                                                           & C1                                                                     & 4/24                                                                          & 7.47     & 5.60            & 5.08                        \\
                                                                             &                                                                        & 16                                                                            & -        & 5.53            & 5.00                        \\
\ding{52}                                                                           & C2                                                                     & 4/24                                                                          & 7.07     & 5.70            & 5.26                        \\
                                                                             &                                                                        & 16                                                                            & -        & 5.70            & 5.20                        \\
\ding{52}                                                                           & C3                                                                     & 4/24                                                                          & 7.07     & 5.68            & 5.21                        \\
                                                                             &                                                                        & 16                                                                            & -        & 5.65            & 5.16                        \\
\bottomrule
\end{tabular}}
\end{table}

\subsection{Ablation study}
Table~\ref{table:ctc} reports the impact of \textit{joint training} with non-streaming loss and different CTC decoder sharing methods on performance.

\textbf{Joint training}. From Table~\ref{table:ctc}, we see that using \textit{joint training} with the non-streaming loss leads to better performance than that without the non-streaming loss. Moreover, according to the analysis in Section~\ref{sec:dual-conv}, the increased amount of parameters is acceptable.

\textbf{CTC decoder}. When using \textit{joint training} with non-streaming, we compared three different CTC decoder sharing methods: (i) \textbf{C1}: A CTC decoder is shared by the output of the top layers in the encoder with the non-streaming mode and the output of the top layers in the encoder with the streaming mode, (ii) \textbf{C2}: A CTC decoder is shared by the output of the top layers in the encoder with the non-streaming mode and the output of the bottom layers in the encoder with the streaming mode and (iii) \textbf{C3}: all outputs of the encoder share a CTC decoder.

As listed in Table~\ref{table:ctc}, the best CER after rescoring could be achieved by using the \textbf{C1}. However, the performance of the greedy search in the bottom layers of its encoder degrades slightly, compared to the other two methods.

Surprisingly, we find that using the full 12-layers of the encoder with a chunk size of 16, the CERs after the greedy search and rescoring are 5.53\% and 5.00\% respectively. That is to say, compared to U2++, fast-U2++ achieves a relative 5.06\% (5.53\% v.s. 5.81\%) CER reduction using the CTC greedy search, which shows the advantage of the \textit{joint training} with the non-streaming loss.

\section{Conclusion}
In this paper, we focus on reducing the partial latency of U2++, which leads to a new method, named fast-U2++. Fast-U2++ outputs partial results from the bottom layers of its encoder with a small chunk. It further uses \textit{knowledge distillation} to improve the token emission latency. Experiments and ablation studies show that fast-U2++ not only significantly reduces the token emission latency, but also achieves similar performance with U2++.

\small
\bibliographystyle{IEEEbib}
\bibliography{myref}

\begin{thebibliography}{10}

\bibitem{li2020developing}
Jinyu Li, Rui Zhao, Zhong Meng, Yanqing Liu, Wenning Wei, Sarangarajan
  Parthasarathy, Vadim Mazalov, Zhenghao Wang, Lei He, Sheng Zhao, et~al.,
\newblock ``Developing rnn-t models surpassing high-performance hybrid models
  with customization capability,''
\newblock in {\em Proc. Interspeech}, 2020.

\bibitem{sainath2020streaming}
Tara~N Sainath, Yanzhang He, Bo~Li, Arun Narayanan, Ruoming Pang, Antoine
  Bruguier, Shuo-yiin Chang, Wei Li, Raziel Alvarez, Zhifeng Chen, et~al.,
\newblock ``A streaming on-device end-to-end model surpassing server-side
  conventional model quality and latency,''
\newblock in {\em ICASSP}. IEEE, 2020, pp. 6059--6063.

\bibitem{graves2006connectionist}
Alex Graves, Santiago Fern{\'a}ndez, Faustino Gomez, and J{\"u}rgen
  Schmidhuber,
\newblock ``Connectionist temporal classification: labelling unsegmented
  sequence data with recurrent neural networks,''
\newblock in {\em Proc. ICML}, 2006, pp. 369--376.

\bibitem{amodei2016deep}
Dario Amodei, Sundaram Ananthanarayanan, Rishita Anubhai, Jingliang Bai, Eric
  Battenberg, Carl Case, Jared Casper, Bryan Catanzaro, Qiang Cheng, Guoliang
  Chen, et~al.,
\newblock ``Deep speech 2: End-to-end speech recognition in english and
  mandarin,''
\newblock in {\em ICML}. PMLR, 2016, pp. 173--182.

\bibitem{graves2012sequence}
Alex Graves,
\newblock ``Sequence transduction with recurrent neural networks,''
\newblock {\em arXiv preprint arXiv:1211.3711}, 2012.

\bibitem{graves2013speech}
Alex Graves, Abdel-rahman Mohamed, and Geoffrey Hinton,
\newblock ``Speech recognition with deep recurrent neural networks,''
\newblock in {\em ICASSP}. IEEE, 2013, pp. 6645--6649.

\bibitem{chan2016listen}
William Chan, Navdeep Jaitly, Quoc Le, and Oriol Vinyals,
\newblock ``Listen, attend and spell: A neural network for large vocabulary
  conversational speech recognition,''
\newblock in {\em ICASSP}. IEEE, 2016, pp. 4960--4964.

\bibitem{chorowski2015attention}
Jan~K Chorowski, Dzmitry Bahdanau, Dmitriy Serdyuk, Kyunghyun Cho, and Yoshua
  Bengio,
\newblock ``Attention-based models for speech recognition,''
\newblock {\em Advances in neural information processing systems}, vol. 28,
  2015.

\bibitem{kim2017joint}
Suyoun Kim, Takaaki Hori, and Shinji Watanabe,
\newblock ``Joint ctc-attention based end-to-end speech recognition using
  multi-task learning,''
\newblock in {\em ICASSP}. IEEE, 2017, pp. 4835--4839.

\bibitem{wu2021u2++}
Di~Wu, Binbin Zhang, Chao Yang, Zhendong Peng, Wenjing Xia, Xiaoyu Chen, and
  Xin Lei,
\newblock ``U2++: Unified two-pass bidirectional end-to-end model for speech
  recognition,''
\newblock {\em arXiv preprint arXiv:2106.05642}, 2021.

\bibitem{dong2018speech}
Linhao Dong, Shuang Xu, and Bo~Xu,
\newblock ``Speech-transformer: a no-recurrence sequence-to-sequence model for
  speech recognition,''
\newblock in {\em ICASSP}. IEEE, 2018, pp. 5884--5888.

\bibitem{gulati2020conformer}
Anmol Gulati, James Qin, Chung-Cheng Chiu, Niki Parmar, Yu~Zhang, Jiahui Yu,
  Wei Han, Shibo Wang, Zhengdong Zhang, Yonghui Wu, et~al.,
\newblock ``Conformer: Convolution-augmented transformer for speech
  recognition,''
\newblock {\em Interspeech}, 2020.

\bibitem{yu2020dual}
Jiahui Yu, Wei Han, Anmol Gulati, Chung-Cheng Chiu, Bo~Li, Tara~N Sainath,
  Yonghui Wu, and Ruoming Pang,
\newblock ``Dual-mode asr: Unify and improve streaming asr with full-context
  modeling,''
\newblock {\em ICLR}, 2021.

\bibitem{narayanan2021cascaded}
Arun Narayanan, Tara~N Sainath, Ruoming Pang, Jiahui Yu, Chung-Cheng Chiu,
  Rohit Prabhavalkar, Ehsan Variani, and Trevor Strohman,
\newblock ``Cascaded encoders for unifying streaming and non-streaming asr,''
\newblock in {\em ICASSP}. IEEE, 2021, pp. 5629--5633.

\bibitem{moritz2021dual}
Niko Moritz, Takaaki Hori, and Jonathan~Le Roux,
\newblock ``Dual causal/non-causal self-attention for streaming end-to-end
  speech recognition,''
\newblock {\em Interspeech}, 2021.

\bibitem{zhang2020unified}
Binbin Zhang, Di~Wu, Zhuoyuan Yao, Xiong Wang, Fan Yu, Chao Yang, Liyong Guo,
  Yaguang Hu, Lei Xie, and Xin Lei,
\newblock ``Unified streaming and non-streaming two-pass end-to-end model for
  speech recognition,''
\newblock {\em arXiv preprint arXiv:2012.05481}, 2020.

\bibitem{yao2021wenet}
Zhuoyuan Yao, Di~Wu, Xiong Wang, Binbin Zhang, Fan Yu, Chao Yang, Zhendong
  Peng, Xiaoyu Chen, Lei Xie, and Xin Lei,
\newblock ``Wenet: Production oriented streaming and non-streaming end-to-end
  speech recognition toolkit,''
\newblock in {\em Proc. Interspeech}, Brno, Czech Republic, 2021, IEEE.

\bibitem{zhang2022wenet}
Binbin Zhang, Di~Wu, Zhendong Peng, Xingchen Song, Zhuoyuan Yao, Hang Lv, Lei
  Xie, Chao Yang, Fuping Pan, and Jianwei Niu,
\newblock ``Wenet 2.0: More productive end-to-end speech recognition toolkit,''
\newblock {\em Interspeech}, 2022.

\bibitem{shangguan2021dissecting}
Yuan Shangguan, Rohit Prabhavalkar, Hang Su, Jay Mahadeokar, Yangyang Shi,
  Jiatong Zhou, Chunyang Wu, Duc Le, Ozlem Kalinli, Christian Fuegen, et~al.,
\newblock ``Dissecting user-perceived latency of on-device e2e speech
  recognition,''
\newblock {\em Interspeech}, 2021.

\bibitem{hinton2015distilling}
Geoffrey Hinton, Oriol Vinyals, Jeff Dean, et~al.,
\newblock ``Distilling the knowledge in a neural network,''
\newblock {\em arXiv preprint arXiv:1503.02531}, vol. 2, no. 7, 2015.

\bibitem{tripathi2020transformer}
Anshuman Tripathi, Jaeyoung Kim, Qian Zhang, Han Lu, and Hasim Sak,
\newblock ``Transformer transducer: One model unifying streaming and
  non-streaming speech recognition,''
\newblock {\em arXiv preprint arXiv:2010.03192}, 2020.

\bibitem{park2019specaugment}
Daniel~S Park, William Chan, Yu~Zhang, Chung-Cheng Chiu, Barret Zoph, Ekin~D
  Cubuk, and Quoc~V Le,
\newblock ``Specaugment: A simple data augmentation method for automatic speech
  recognition,''
\newblock {\em Interspeech}, 2019.

\bibitem{chang2020low}
Shuo-Yiin Chang, Bo~Li, David Rybach, Yanzhang He, Wei Li, Tara~N Sainath, and
  Trevor Strohman,
\newblock ``Low latency speech recognition using end-to-end prefetching.,''
\newblock in {\em Interspeech}, 2020, pp. 1962--1966.

\bibitem{tian2020synchronous}
Zhengkun Tian, Jiangyan Yi, Ye~Bai, Jianhua Tao, Shuai Zhang, and Zhengqi Wen,
\newblock ``Synchronous transformers for end-to-end speech recognition,''
\newblock in {\em ICASSP}. IEEE, 2020, pp. 7884--7888.

\bibitem{zhang2020streaming}
Shiliang Zhang, Zhifu Gao, Haoneng Luo, Ming Lei, Jie Gao, Zhijie Yan, and Lei
  Xie,
\newblock ``Streaming chunk-aware multihead attention for online end-to-end
  speech recognition,''
\newblock {\em Interspeech}, 2020.

\bibitem{inaguma2020enhancing}
Hirofumi Inaguma, Masato Mimura, and Tatsuya Kawahara,
\newblock ``Enhancing monotonic multihead attention for streaming asr,''
\newblock {\em Interspeech}, 2020.

\bibitem{wang2021wnars}
Zhichao Wang, Wenwen Yang, Pan Zhou, and Wei Chen,
\newblock ``Wnars: Wfst based non-autoregressive streaming end-to-end speech
  recognition,''
\newblock {\em arXiv preprint arXiv:2104.03587}, 2021.

\end{thebibliography}

\end{document}